# Beyond Memristor: Neuromorphic Computing Using Meminductor

## Frank Z. Wang

Division of Computing, Engineering & Mathematics, University of Kent, UK (f.z.wang@kent.ac.uk)

*Abstract—* **Memristor (resistor-with-memory), inductor-with-memory (meminductor) and capacitor-with-memory (memcapacitor) have different roles to play in novel computing architectures. We found that a coil with a magnetic core is an inductor with memory (meminductor) in terms of its inductance $L(q)$ being a function of the charge $q$. The history of the current passing through the coil is remembered by the magnetization inside the magnetic core. Such a meminductor can play a unique role (that cannot be played by a memristor) in neuromorphic computing, deep learning and brain-inspired since the time constant ($t_0 = \sqrt{LC}$) of a neuromorphic *RLC* circuit is jointly determined by the inductance $L$ and capacitance $C$, rather than the resistance $R$. As an experimental verification, this newly invented meminductor was used to reproduce the observed biological behaviour of amoebae (the memorizing, timing and anticipating mechanisms). In conclusion, a beyond-memristor computing paradigm is theoretically sensible and experimentally practical.**

*Keywords—* memristor, meminductor, novel computing architectures, non-Turing machine, neuromorphic computing, deep learning, brain-inspired computing.

## I. INTRODUCTION

Memristor is an ideal candidate for non-Turing machines due to its compact processing-in-memory architecture. As a sister of memristor (resistor-with-memory), inductor-with-memory (meminductor) has a unique role to play in neuromorphic computing systems, novel computing architectures and dynamical neural networks.

An inductor, typically consisting of an insulated wire wound into a coil, stores energy in a magnetic flux $\varphi$ surrounding it when a current $i$ flows through it. When the current changes, the time-varying magnetic flux induces a voltage across the coil, described by Faraday's law [1]. Such an inductor is characterized by its inductance $L = \frac{\varphi}{i}$. In SI, the unit of inductance is the henry (*H*). As shown in Fig. 1, by adding a magnetic core made of a ferromagnetic material such as iron inside the coil, the magnetizing flux from the coil induces magnetization in the material, increasing the magnetic flux. The high permeability of a ferromagnetic core can increase the inductance of a coil by a factor of several thousand over what it would be without it [1].

Organisms like amoebae exhibit primitive learning and the (memorizing, timing and anticipating) mechanism. Their adaptive behavior was emulated by a memristor-based *RLC* circuit [2]. Motivated by this work, we will design a meminductor-based neuromorphic architecture that self-adjusts its inherent resonant frequency in a natural way following the external stimuli frequency. In contrast to the previous work, our innovation is that this architecture uses a unique meminductor to increment its time constant and subsequently decrement its resonant frequency to match the stimuli frequency. It is our intention to use this architecture to help better investigate the cellular origins of primitive intelligence. This is also the significance of this sort of research in terms of not only understanding the primitive learning but also developing a novel computing architecture.

In this article, we first prove that a coil structure with a magnetic core is actually a meminductor since its inductance is not a constant any longer and then experimentally verify this new device in neuromorphic computing.

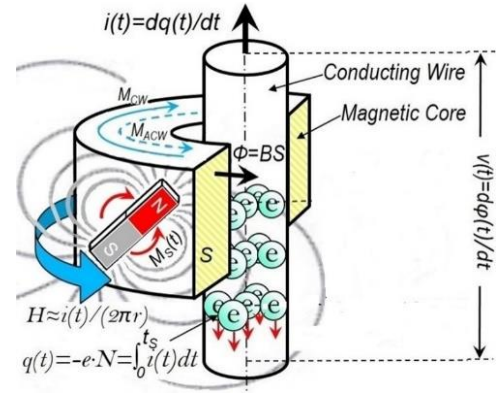

Fig. 1. In this study, we found that a coil with a magnetic core is actually an inductor with memory (meminductor) in terms of its inductance being a function of the charge. The Oersted field generated by the current $i$ rotates or switches the magnetization $M$ inside the core and consequently the switched flux $\varphi$ induces a voltage v across the conductor. The history of the current passing through the coil [$\int i(t)dt = q(t)$] is remembered by the magnetization inside the magnetic core.

## II. LLG MODEL FOR THE COIL-CORE STRUCTURE

Next, we produce a theory to physically describe the current-flux interaction in a conducting coil with a magnetic core. For the sake of convenience, the magnetic core is assumed to be a single-domain cylinder with uniaxial anisotropy in the approximate sense: the magnetization is uniform and rotates in unison [3]. In an ideal case, there is a negligible amount of eddy current damping and parasitic "capacitor" effect.

It was found that the rotational process dominates the fast reversal of square loop ferrites with a switching coefficient $S_w = 0.2\ Oe \cdot \mu s$ [4]. The rotational model for the coil-core structure is shown in Fig. 2.

The Landau–Lifshitz–Gilbert equation [5][6] is:

$$(1+g^2)\frac{d\vec{M_S}(t)}{dt} = -|\gamma|[\vec{M_S}(t) \times \vec{H}] - \frac{g|\gamma|}{M_S}[\vec{M_S}(t) \times (\vec{M_S}(t) \times \vec{H})],$$

where $g$ is the damping factor and $\Upsilon$ is the gyromagnetic ratio.

The 1st term of the right-hand side can be rewritten as: $-|\gamma|\vec{M_S}(t) \times \vec{H} = -|\gamma|(M_S sin\theta sin\psi H\vec{\imath} - M_S sin\theta cos\psi H\vec{\jmath})$. This term has no $\vec{k}$ component (along Z) and does not contribute to $M_Z$.

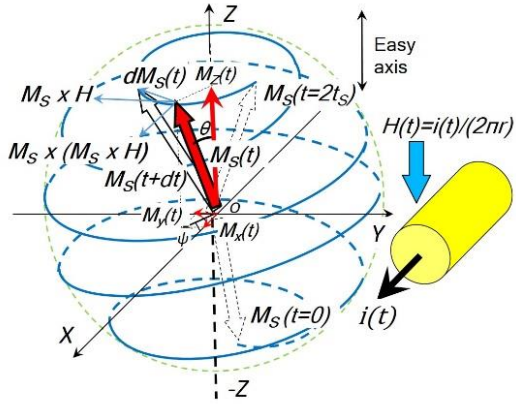

Fig. 2. The rotational model used in the coil-core structure. If the magnetic field $\vec{H}$ is applied in the Z direction, the saturation magnetization vector $\vec{M_S}(t)$ follows a precession trajectory (blue) from its initial position ($\theta_0 \approx \pi$, $m_0 \approx -1$) and the angle $\theta$ decreases with time continuously until ($\theta \approx 0$, $m \approx 1$), i.e., the magnetization $\vec{M_S}(t)$ reverses itself and is eventually aligned with the magnetic field $\vec{H}$.

The 2nd term can be rewritten as:

$$-\frac{g|\gamma|}{M_S}[\vec{M_S}(t) \times (\vec{M_S}(t) \times \vec{H})] =$$

$$= -\frac{g|\gamma|}{M_S}(M_S sin\theta cos\psi \vec{\imath} + M_S sin\theta sin\psi \vec{\jmath}$$

$$+ M_S cos\theta \vec{k}) \times [M_S sin\theta sin\psi H\vec{\imath} - M_S sin\theta cos\psi H\vec{\jmath}]$$

$$= -\frac{g|\gamma|}{M_S}[-M_S sin\theta cos\psi M_S sin\theta cos\psi H - M_S sin\theta sin\psi M_S sin\theta sin\psi H]\vec{k}$$

$$= g|\gamma|M_S H[sin^2\theta cos^2\psi + sin^2\theta sin^2\psi]\vec{k} = g|\gamma|M_S H sin^2\theta \vec{k}$$

$$= g|\gamma|M_S H(1 - cos^2\theta)\vec{k} = g|\gamma|M_S H[1 - \left(\frac{M_Z}{M_S}\right)^2]\vec{k}.$$

From the above, we can obtain the following equation:

$$(1+g^2)\frac{dM_Z(t)}{dt} = g|\gamma|M_S H[1 - \left(\frac{M_Z}{M_S}\right)^2]. \quad (1)$$

Assuming $m(t) = \frac{M_Z(t)}{M_S}$, we can obtain

$$\frac{dm(t)}{dt} = \frac{g|\gamma|H}{(1+g^2)}[1 - m^2(t)] = \frac{1}{S_W}i(t)[1 - m^2(t)], \quad (2)$$

The threshold for magnetization switching is automatically taken into account because the switching coefficient is defined based on the threshold field $H_0$, which is one to two times the coercive force $H_C$ [7][8][9].

The hyperbolic function $tanh$ has $\frac{d}{dx}tanh\,x = 1 - tanh^2 x$ and the derivative of a function of function has $\frac{du}{dx} = \frac{du}{dy}\frac{dy}{dx}$; therefore, it is reasonable to assume that

$$m(t) = tanh\left[\frac{q(t)}{S_W} + C\right], \quad (3)$$

where $\frac{d}{dt}q(t) = i(t)$ and $C$ is a constant of integration such that $C = tanh^{-1}m_0$ if $q(t=0)=0$ (assuming the charge does not accumulates at any point) and $m_0$ is the initial value of $m$.

$dM_Z/dt$ can be observed by the voltage $v(t)$ induced:

$$\mu_0 S \frac{dM_Z}{dt} = S\frac{dB_Z}{dt} = \frac{d\varphi_Z}{dt} = -v(t), \quad (4)$$

where $\mu_0$ is the permeability and $S$ is the cross-sectional area.

Equation (4) results in

$$\varphi = \mu_0 SM + C' = \mu_0 SM_S m + C', \quad (5)$$

where $C'$ is another constant of integration.

Combining Eq. (3) and Eq. (5) and assuming $\varphi(t=0) = 0$, we have $C' = -\mu_0 SM_S m_0$, so

$$\varphi = \mu_0 SM_s\left[tanh\left(\frac{q}{S_W} + tanh^{-1}m_0\right) - m_0\right]. \quad (6)$$

Beyond the first-order setting, a second-order circuit element such as a meminductor requires double-time integrals of voltage and current, namely, $\sigma = \int q dt = \iint i dt$ and $\rho = \int \varphi dt = \iint v dt$. With the use of these additional variables [9][10], we accommodate a meminductor, a mem-capacitor and other second-order circuit elements with memory. By integrating Eq. (6), we have:

$$\rho = \int_{\tau=-\infty}^{t} \varphi\,d\tau = \mu_0 SM_s \int_{\tau=-\infty}^{t}\left[tanh\left(\frac{q}{S_W} + tanh^{-1}m_0\right) - m_0\right] d\tau. \quad (7)$$

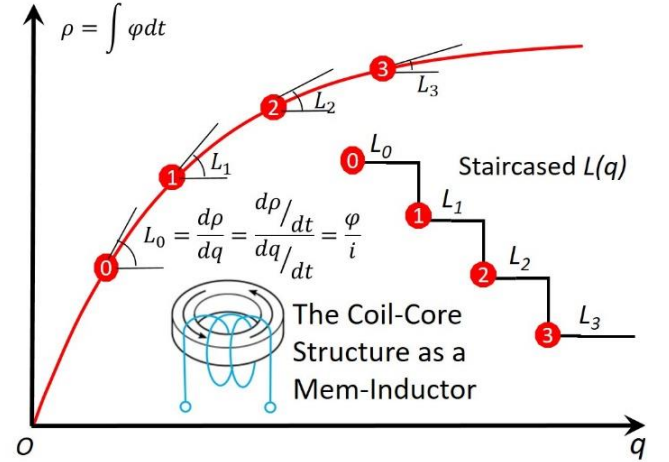

Fig. 3. The constitutional $\rho$-$q$ curve of the meminductor. It complies with the three criteria for the ideality of an ideal circuit element with memory [11][12]: a. nonlinear; b. continuously differentiable; and c. strictly monotonically increasing. With the accumulation of the charge, $L(q) = \frac{d\rho}{dq}$ decreases like a staircase.

Since $\int tanh\,x\,dx = ln(cosh\,x) + C$, we have

$$\rho = \mu_0 SM_s\,ln\left\{cosh\left[tanh\left(\frac{q}{S_W} + tanh^{-1}m_0\right) - m_0\right]\right\} + C \triangleq \hat\rho(q). \quad (8)$$

Therefore, we have:

$$L = \frac{\varphi}{i} = \frac{\mu_0 SM_s\left[\tanh\left(\frac{q}{S_W}+\tanh^{-1}m_0\right)-m_0\right]}{dq/dt} \triangleq L(q), \quad (9)$$

where the denominator is still a function of the charge $q = \hat{q}(t)$ since $\frac{dq}{dt} = i(t) = i[\hat{q}^{-1}(q)]$.

Based on Eq. (8), a typical $\rho - q$ curve is depicted in Fig. 3 with $m_0$=-0.964 (this value reflects the intrinsic fluctuation; otherwise, $M$ reverts to the stable equilibria $m_0 = \pm 1$).

### III. EXPERIMENTAL VERIFICATION OF THE ROTATIONAL MODEL

To verify the validity/accuracy of the above rotational model, Eq. (3) with $H(t) \propto i(t)$ is used to reproduce various $m$-$H$ loops in Fig. 4.

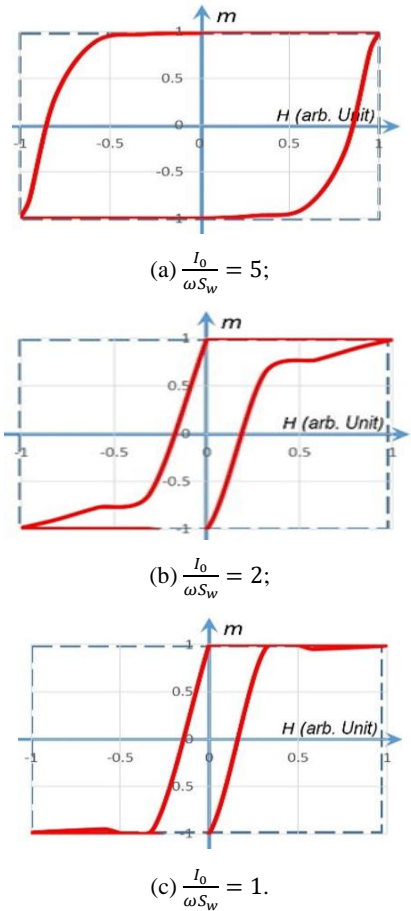

(a) $\frac{I_0}{\omega S_w} = 5$;

(b) $\frac{I_0}{\omega S_w} = 2$;

(c) $\frac{I_0}{\omega S_w} = 1$.

Fig. 4. The $m$-$H$ hysteresis loops simulated by the rotational model. The solid line in red represents a gradual $m(t)$ rotation (with a finite slope) under $H(t) \propto i(t) = I_0 \sin \omega t, m_0 = \pm 0.99$. The dashed line in blue represents a fast $m(t)$ rotation (with an infinite slope) under a step-function $H$.

As a comparison, a typical $m$-$H$ loop of real-world magnetic materials is displayed in Fig. 5. The above simulations clearly validate Cushman's conclusion that "the rotational model is applicable to the driving current of an arbitrary waveform" [7].

As another comparison, a simulated loop based on $m = \tanh(A * (H \pm H_C))]$ is displayed in Fig. 6. The equivalence of formula $m = \tanh(A * (H \pm H_C))]$ and formula $m(t) = \tanh\left[\frac{1}{S_W}(q(t) \pm S_W \tanh^{-1}|m_0|)\right]$ indicates that the rotational model is good enough to reproduce a sine-wave response.

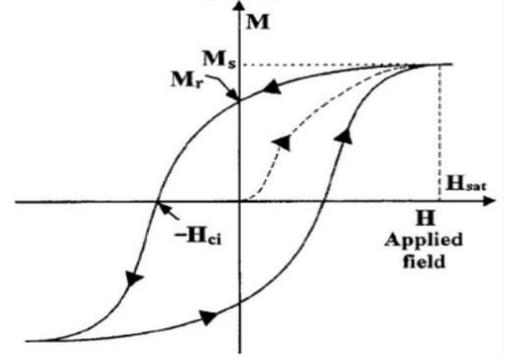

Fig. 5. A typical $m$-$H$ loop of real-world magnetic materials [13].

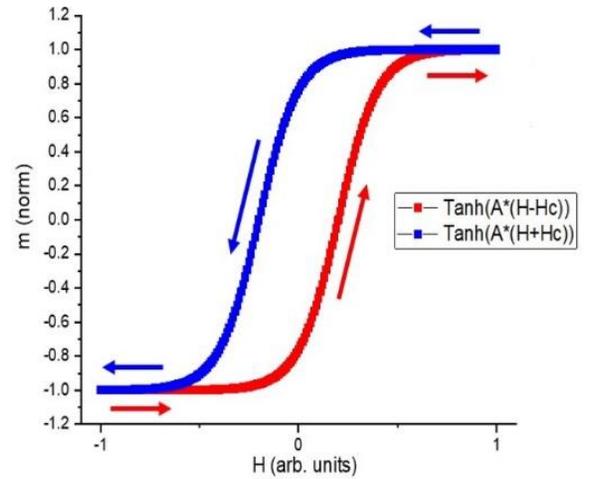

Fig. 6. The simulated $m$-$H$ loop based on $m = \tanh(A * (H \pm H_C))]$ ($H_C$ is the coercive force) with a sine-wave input current. Two $\tanh$ values are used, and a horizontal shift is applied to each branch to obtain hysteresis.

### IV. SIMULATIONS & EXPERIMENTS OF A COIL-CORE MEM-INDUCTOR FOR NEUROMORPHIC COMPUTING

Nature exhibits unconventional ways of processing information. Taking amoebae as an example, they display memorizing, timing and anticipating mechanisms, which may represent the origins of primitive learning. A circuit element with memory can be used to mimic these behaviours in terms of being plastic according to the dynamic history [14][15][16].

As shown in Fig. 7, a simple $RLC$ neuromorphic circuit using a coil-core meminductor, $L(q)$, is designed. The temperature controlling the motion of an amoeba is analogous to the input voltage, $V_{in}$, whereas the output voltage, $V_{out}$, is analogous to the locomotive speed of the amoeba.

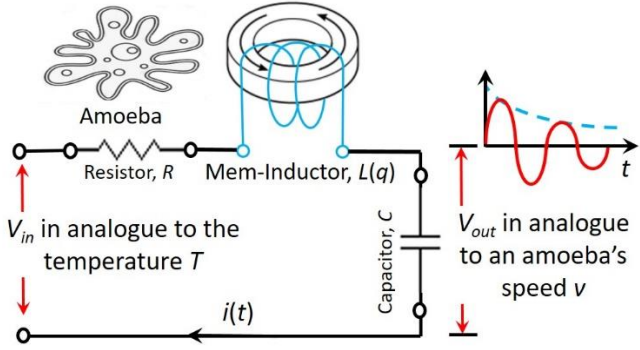

Fig. 7. An *RLC* neuromorphic circuit using a coil-core meminductor, $L(q)$, to scan a frequency range. An amoeba's behaviour is simulated with the damped oscillations of this circuit.

With the progress of time, the circuit's resonance frequency automatically scans the following frequency range:

$$f_0 = \frac{1}{2\pi\sqrt{L(q)C}} = \frac{1}{2\pi\sqrt{L(\int i(t)dt)C}} \quad (10)$$

When the ramping circuit resonance frequency, $f_0$, hits the (temperature) stimulus frequency, $f_{sti}$, at a time point, a resonance is triggered.

This neuromorphic circuit in Fig. 7 using a coil-core meminductor reasonably reproduces a behaviour that was observed on amoebae: in response to the input stimulus pulses (representing the temperature drops), the circuit reduces the amplitude of its output (representing the amoeba's speed) at the corresponding time points. As demonstrated in Fig. 8, long-lasting responses for spontaneous in-phase slow down (SPS) [14][15] are both simulated and experimented: the amoeba being exposed to the three temperature drops slows down or even stops at the corresponding time points $S_1$, $S_2$ and $S_3$. Remarkably, the amoeba is found to slow down even if the temperature drops do not occur at $C_1$, $C_2$ and $C_3$ (that are naturally anticipated by the amoeba after the three consecutive drops are experienced at $S_1$, $S_2$ and $S_3$).

The experimental setup of the neuromorphic circuit in Fig. 8 is as below: $L[q(t)] = L[\int i(t)dt]$ starts at 2 $H$ and then decreases by 20% after each stimulus pulse. The circuit's resonance frequency, determined by the staircased $L(q)$ (Fig. 3), increases itself with the increased number of oncoming stimulus pulses. This simulation in Fig. 8(a) agrees with our experiment in Fig. 8(b) on a hardware emulator built with a dsPIC30F2011 microcontroller, an MCP4261 digital potentiometer, and a differential 12-bit ADC converter [16].

This experiment vividly demonstrates Amoebae's three mechanisms: 1. the memorizing mechanism (the amoeba remembers the three temperature drops at $S_1$, $S_2$ and $S_3$); 2. the timing mechanism (the amoeba slows down at the correct time points $C_1$, $C_2$ and $C_3$ despite no temperature drops at these time points); and 3. the anticipating mechanism (why the amoeba slows down actively is because it anticipates the future possible drops at $C_1$, $C_2$ and $C_3$ based on its memory of $S_1$, $S_2$ and $S_3$ although these temperature drops at $C_1$, $C_2$ and $C_3$ do not occur). Remarkably, these memorizing/timing/anticipating mechanisms are implemented by our newly invented coil-core meminductor in terms of using the magnetization to remember the current history, adapting automatically the time constant determined by $L(q)$ to the stimulus and triggering the resonance, respectively.

This neuromorphic circuit is a deep learning neural network [17] with multiple layers between the input and output layers, as shown in Fig. 9. The meminductor $L(q)$ and capacitor $C$ store energy in the form of magnetic flux and electric field, respectively, whereas resistor $R$ only consumes energy. Energy can be transferred from one form to the other, which is oscillatory with a resonance frequency ($f_0 = \frac{1}{2\pi\sqrt{L(q)C}}$). The resistance $R$ dampens the oscillation, diminishing it with time. Not strictly speaking, such a damped oscillation may be vividly approximated by $e^{-\alpha t}\sin 2\pi f_0 t$, where $\alpha = \frac{R}{2L(q)}$ is the damping factor.

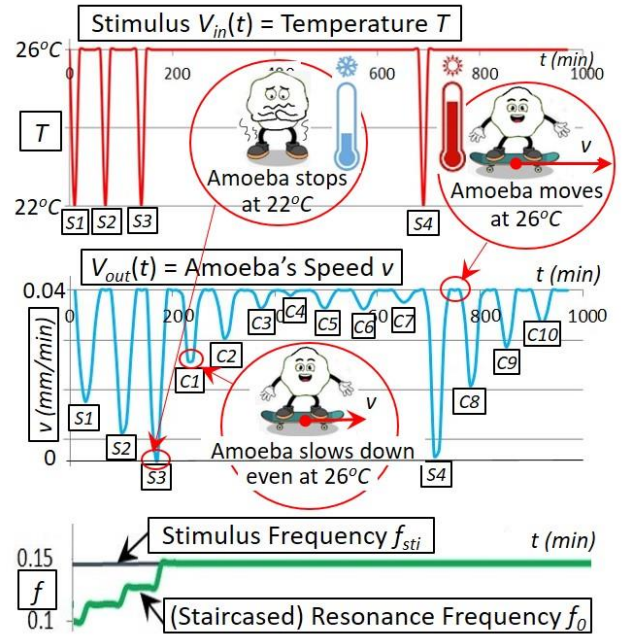

(a) Simulated response (halved) of the neuromorphic *RLC* circuit;

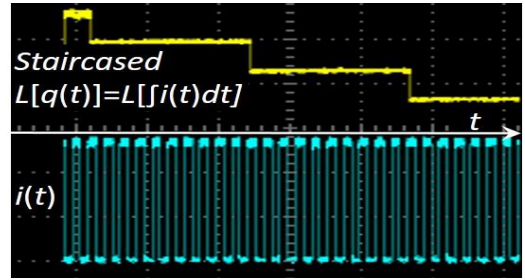

(b) Measured (staircased) inductance on a hardware emulator.

Fig. 8. Simulated and experimental responses of the neuromorphic circuit. $L[q(t)] = L[\int i(t)dt]$ starts at 2 $H$ and then decreases by 20% after each stimulus pulse. The circuit's resonance frequency, determined by the staircased $L(q)$ (Fig. 3), increases itself with the increased number of oncoming stimulus pulses. This simulation in (a) agrees with our experiment in (b) on a hardware emulator built with a dsPIC30F2011 microcontroller, an MCP4261 digital potentiometer, and a differential 12-bit ADC converter.

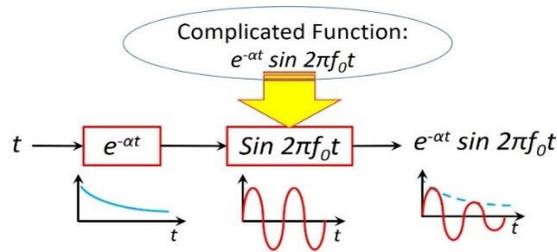

Fig. 9. The neuromorphic *RLC* circuit in Fig. 7 is a deep learning neural network with multiple layers between the input and output layers. The complicated function $e^{-\alpha t}\sin 2\pi f_0 t$ is decomposed into two simple functions: $e^{-\alpha t}$ and $\sin 2\pi f_0 t$, each of which can be implemented in one layer. The former is determined by *R* and *L(q),* whereas the latter is determined by *L(q)* and *C*.

## V. CONCLUSIONS & DISCUSSIONS

Memristor (resistor-with-memory), inductor-with-memory (meminductor) and capacitor-with-memory (memcapacitor) have different roles to play in neuromorphic computing systems, novel computing architectures and dynamical neural networks. In this study, we found that a coil with a magnetic core is actually an inductor with memory (meminductor) in terms of its inductance being a function of the charge. This meminductor can play a unique role (that cannot be played by a memristor) in neuromorphic computing [18][19], deep learning [17] and brain-inspired computing [20][21][22] since the time constant ($t_0 = \sqrt{LC}$) of a neuromorphic *RLC* circuit is jointly determined by the inductance $L$ and capacitance $C$, rather than the resistance $R$. As an experimental verification, this new meminductor was used to reasonably reproduce the observed biological behaviour of amoebae, in which the resonance frequency tracks the stimulus frequency. In conclusion, a beyond-memristor computing paradigm is theoretically sensible and experimentally practical.

Nature exhibits unconventional ways of storing and processing information and circuit elements with memory mimic the dynamical behaviours of some biological systems in terms of being plastic according to the history of the systems. As a practical application, the Pavlovian experiment on conditioned reflex is reproduced by a memristor neural network with the aid of the so-called "delayed switching" effect [23][24]. In this application, the total length of the stimuli sequence, the frequency of the stimuli sequence and the spike width are carefully adjusted such that the time delay point of the memristor synapse should not be exceeded while only one neuron fires. In many applications, it is not feasible and possible to solve the problems with conventional computational models and methods (i.e. the Turing machine [25][26][27][28], the von Neumann architecture [29][30][31][32]). As demonstrated above, neuromorphic architectures may help.

Understanding the brain with non-linear dynamics and extreme complexity is still a great challenge since the human brain has $10^{11}$ neurons and $10^{14}$ synapses (each neuron is connected to up to 20,000 synapses) [33][34][35][36][37]. By coincidence, as one of the simplest creatures or organisms existing on earth, unicellular amoebae display some mysterious brain-like behaviours in terms of controlling their actions [38][39][40][41][42]. Their memorizing, timing and anticipating mechanisms may represent the origins of primitive learning.

Evolution of life includes the process of evolving intelligence in charge of controlling and predicting their behaviour. In 1952, Hodgkin and Huxley developed an equivalent circuit to explain the initiation/propagation of action potentials and the underlying ionic mechanisms in the squid giant axon [18][43][44][45][46][47]. They were awarded the Nobel Prize in Physiology or Medicine for this work in 1963. In the so-called Hodgkin–Huxley model, an electrical circuit representing each cell consists of a linear resistor, a capacitor, three batteries, and two unconventional elements identified by Hodgkin and Huxley as time-varying resistors. In 2012, these two potassium and sodium time-varying resistors was substituted by a potassium ion-channel memristor, and a sodium ion-channel memristor, respectively [19][20]. This is a great progress on neural physiology and brain science for over 70 years in terms of exploring the origins of primitive learning from an evolutionary perspective.

In this work, we developed a meminductor-based neuromorphic architecture that self-adjusts its inherent resonant frequency in a natural way following the external stimuli frequency. In contrast to the previous work, our innovation is that this architecture uses a unique meminductor to increment its time constant and subsequently decrement its resonant frequency to match the stimuli frequency. This architecture may help better investigate the cellular origins of primitive intelligence [48][49][50]. This sort of research is significant in terms of not only understanding the primitive learning but also developing a novel computing architecture, which will be much more integrated with our physical and social environment, capable of self-learning and capable of process and distribute big data at an unprecedented scale [51][52]. This will require new deigns, new theories, new paradigms, and close interactions with application experts in the sense that new bio-inspired (neurosynaptic) and non-Turing-inspired computing platforms are moving away from traditional computer architecture design [52].

**Acknowledgment**
This research was partially funded by an EC grant "Re-discover a periodic table of elementary circuit elements", PIIFGA2012332059, Marie Curie Fellow: Leon Chua (UC Berkeley), Scientist-in-charge: Frank Wang (University of Kent).

**References**
[1] C. Alexander, M. Sadiku, Fundamentals of Electric Circuits (3 ed.). McGraw-Hill. p. 211.
[2] Y. V. Pershin, S. La Fontaine and M. Di Ventra, Memristive model of amoeba learning, PHYSICAL REVIEW E 80, 021926, 2009
[3] M. J. Donahue, D. G. Porter, "Analysis of switching in uniformly magnetized bodies". IEEE Transactions on Magnetics. V.38, I.5, pp.2468-2470, Dec., 2002, 10.1109/TMAG.2002.803616.

[4] E. M. Gyorgy, "Rotational model of flux reversal in square loop ferritcs" J. Appl. Phys., V.28, I.9, pp. 1011-1015, Sep., 1957, 10.1063/1.1722897.

[5] L.D. Landau, E.M. Lifshitz, "Theory of the dispersion of magnetic permeability in ferromagnetic bodies". Phys. Z. Sowjetunion. 8, 153, 1935.

[6] T.L. Gilbert, "A Lagrangian formulation of the gyromagnetic equation of the magnetic field". Physical Review. 100: 1243, 1955.

[7] E. M. Gyorgy, "Rotational model of flux reversal in square loop ferritcs" J. Appl. Phys., V.28, I.9, pp. 1011-1015, Sep., 1957, 10.1063/1.1722897.

[8] N. Cushman, "Characterization of Magnetic Switch Cores", IRE Transactions on Component Parts, V.8, I.2, pp.45-50, Jun., 1961, 10.1109/TCP.1961.1136600.

[9] N. Menyuk, J. Goodenough, "Magnetic materials for digital computer components I", J. Appl. Phys., V.26, I.1, Aug., 1955, 10.1063/1.1721867.

[10] R. Riaza, "Second order mem-circuits", International Journal of Circuit Theory and Application, http://mc.manuscriptcentral.com/ijcta, accessed in January 2021.

[11] L. Chua, "Memristor—The Missing Circuit Element", IEEE Transactions on Circuit Theory CT-18(5), 1971: 507-519.

[12] P.S.Georgiou, M. Barahona, S. N.Yaliraki, E. M.Drakakis, "On memristor ideality and reciprocity", Microelectronics Journal, 45 1363–1371, 2014.

[13] C. Rudowicz and H. W. F. Sung, American Journal of Physics 71, 1080 (2003).

[14] Y. Pershin and M. D. Ventra, "Experimental demonstration of associative memory with memristive neural networks", Nature Precedings, 18 May 2009

[15] Y. Pershin and M. D. Ventra, L. Chua, Circuit Elements With Memory: Memristors, Memcapacitors, and Meminductors, https://arxiv.org/pdf/0901.3682.pdf, 2009.

[16] F. Z. Wang, L. O. Chua, X. Yang, N. Helian, R. Tetzlaff, T. Schmidt, L. Li, J. M. Carrasco, W. Chen, D. Chu, "Adaptive Neuromorphic Architecture (ANA)", Special Issue on Neuromorphic Engineering: from Neural Systems to Brain-Like Engineered Systems, Neural Networks, Vol.45, September 2013, pp.111-116, Elsevier, online in March 2013. doi.org/10.1016/j.neunet.2013.02.00.

[17] Y. LeCun, Y. Bengio, G. Hinton, "Deep Learning". Nature. 521 (7553): 436–444. Natur, 2015.

[18] A. Hodgkin and A. Huxley, A quantitative description of membrane current and its application to conduction and excitation in nerve. J. Physiol. 117:500–544. PMID 12991237, 1952

[19] L. Chua, V. Sbitnev, & H. Kim, "Hodgkin–Huxley axon is made of memristors" Int. J. Bifurcation and Chaos 22, 1230011-1–48, 2012.

[20] L. Chua, V. Sbitnev, & H. Kim, "Neurons Are Poised Near The Edge Of Chaos", Int. J. Bifurcation and Chaos, Vol. 22, No. 4, 1250098, 2012.

[21] M. Sah, H. Kim, L. Chua, "Brains Are Made of Memristors", IEEE circuits and systems magazine, First Quarter 2014.

[22] Eric J. Chaisson, Cosmic Evolution – Biological, Havard University Course Syllabus, Version 7, 2012.

[23] Frank Wang, et al., "Delayed switching applied to memristor neural networks", Journal of Applied Physics, DOI: 10.1063/1.3672409 Date: 15-December, 2011.

[24] F.Z. Wang, et al., "Delayed Switching in Memristors and Memristive Systems", IEEE Electron Device Letters, V.31, I.7, 2010.

[25] Minsky 1967:107 "In his 1936 paper, A. M. Turing defined the class of abstract machines that now bear his name. A Turing machine is a finite-state machine associated with a special kind of environment -- its tape -- in which it can store (and later recover) sequences of symbols," also Stone 1972:8 where the word "machine" is in quotation marks.

[26] Stone 1972:8 states "This "machine" is an abstract mathematical model", also cf. Sipser 2006:137ff that describes the "Turing machine model". Rogers 1987 (1967):13 refers to "Turing's characterization", Boolos Burgess and Jeffrey 2002:25 refers to a "specific kind of idealized machine".

[27] Sipser 2006:137 "A Turing machine can do everything that a real computer can do".

[28] Cf. Sipser 2002:137. Also, Rogers 1987 (1967):13 describes "a paper tape of infinite length in both directions". Minsky 1967:118 states "The tape is regarded as infinite in both directions". Boolos Burgess and Jeffrey 2002:25 include the possibility of "there is someone stationed at each end to add extra blank squares as needed".

[29] von Neumann, John (1945), First Draft of a Report on the EDVAC (PDF), archived from the original (PDF) on March 14, 2013, retrieved August 24, 2011.

[30] Markgraf, Joey D. (2007), The Von Neumann Bottleneck, archived from the original on December 12, 2013.

[31] MFTL (My Favorite Toy Language) entry Jargon File 4.4.7, retrieved July 11, 2008.

[32] Turing, Alan M. (1936), "On Computable Numbers, with an Application to the Entscheidungsproblem", Proceedings of the London Mathematical Society, 2 (published 1937), vol. 42, pp. 230–265, doi:10.1112/plms/s2-42.1.230, S2CID 73712 and Turing, Alan M. (1938), "On Computable Numbers, with an Application to the Entscheidungsproblem. A correction", Proceedings of the London Mathematical Society, 2 (published 1937), vol. 43, no. 6, pp. 544–546, doi:10.1112/plms/s2-43.6.544.

[33] "Encephalo- Etymology". Online Etymology Dictionary. Archived from the original on October 2, 2017. Retrieved October 24, 2015.

[34] Parent, A.; Carpenter, M.B. (1995). "Ch. 1". Carpenter's Human Neuroanatomy. Williams & Wilkins. ISBN 978-0-683-06752-1.

[35] Bigos, K.L.; Hariri, A.; Weinberger, D. (2015). Neuroimaging Genetics: Principles and Practices. Oxford University Press. p. 157. ISBN 978-0-19-992022-8.

[36] Cosgrove, K.P.; Mazure, C.M.; Staley, J.K. (2007). "Evolving knowledge of sex differences in brain structure, function, and chemistry". Biol Psychiatry. 62 (8): 847–855. doi:10.1016/j.biopsych.2007.03.001. PMC 2711771. PMID 17544382.

[37] Molina, D. Kimberley; DiMaio, Vincent J.M. (2012). "Normal Organ Weights in Men". The American Journal of Forensic Medicine and Pathology. 33 (4): 368–372. doi:10.1097/PAF.0b013e31823d29ad. ISSN 0195-7910. PMID 22182984. S2CID 32174574.

[38] "Amoeba" Archived 22 November 2015 at the Wayback Machine at Oxforddictionaries.com

[39] Singleton, Paul (2006). Dictionary of Microbiology and Molecular Biology, 3rd Edition, revised. Chichester, UK: John Wiley & Sons. pp. 32. ISBN 978-0-470-03545-0.

[40] David J. Patterson. "Amoebae: Protists Which Move and Feed Using Pseudopodia". Tree of Life web project. Archived from the original on 15 June 2010. Retrieved 21 September 2009.

[41] "The Amoebae". The University of Edinburgh. Archived from the original on 10 June 2009.

[42] Wim van Egmond. "Sun animalcules and amoebas". Microscopy-UK. Archived from the original on 4 November 2005. Retrieved 23 October 2005.

[43] Nelson ME (2005) Electrophysiological Models In: Databasing the Brain: From Data to Knowledge. (S. Koslow and S. Subramaniam, eds.) Wiley, New York, pp. 285–301

[44] Gray DJ, Wu SM (1997). Foundations of cellular neurophysiology (3rd ed.). Cambridge, Massachusetts [u.a.]: MIT Press. ISBN 978-0-262-10053-3.

[45] Krapivin, Vladimir F.; Varotsos, Costas A.; Soldatov, Vladimir Yu. (2015). New Ecoinformatics Tools in Environmental Science : Applications and Decision-making. Springer. pp. 37–38. ISBN 9783319139784.

[46] Rakowski RF, Gadsby DC, De Weer P (May 1989). "Stoichiometry and voltage dependence of the sodium pump in voltage-clamped, internally

dialyzed squid giant axon". The Journal of General Physiology. 93 (5): 903–41. doi:10.1085/jgp.93.5.903. PMC 2216238. PMID 2544655.

[47] Hille B (2001). Ion channels of excitable membranes (3rd ed.). Sunderland, Massachusetts: Sinauer. ISBN 978-0-87893-321-1.

[48] B., F. Primitive Intelligence and Environment. Nature 142, 774 (1938). https://doi.org/10.1038/142774a0

[49] Primitive intelligence, New Scientist, 27 September 2000.

[50] R Näätänen 1 , M Tervaniemi, E Sussman, P Paavilainen, I Winkler, "Primitive intelligence" in the auditory cortex, PMID: 11311381 DOI: 10.1016/s0166-2236(00)01790-2, 2001.

[51] V Kindratenko, Novel Computing Architectures, Computing in Science & Engineering, 2009.

[52] Novel computing platforms and information processing approaches, https://csl.illinois.edu/research/impact-areas/health-it/novel-computing-platforms-and-information-processing-approaches, accessed 6th February 2023.